\documentstyle[aps,preprint,eqsecnum]{revtex}
\newcommand{\be}{\begin{eqnarray}}
\newcommand{\e}{\end{eqnarray}}
\newcommand{\p}{\partial}
\newcommand{\w}{\wedge}
\newcommand{\ep}{\epsilon}
\newcommand{\de}{\delta}
\newcommand{\pp}{\perp}
\begin{document}

\title{Electromagnetic Duality on The Light-Front in The Presence of External
Sources}

\author{Asmita Mukherjee\thanks{e-mail: asmita@tnp.saha.ernet.in} and
Somdatta Bhattacharya\thanks{e-mail: som@tnp.saha.ernet.in} \\
Saha Institute of Nuclear Physics, 1/AF, Bidhannagar,
            Calcutta 700064 India}

\date{April 13, 1999}

\maketitle
\begin{abstract}

We investigate the issue of electromagnetic duality on the light front.
We work with Zwanziger's theory of electric and magnetic sources which is
appropriate for treating duality.  
 When quantized on the light-front in the light front gauge, 
this theory yields 
two independent phase space 
degrees of freedom, namely the two transverse field components,
 the right number to describe the gauge field sector of 
normal light-front QED and also the appropriate
commutator between them. The electromagnetic
 duality transformation formulated in terms of them is similar in form to the
Susskind transformation proposed for the free theory, provided 
one identifies them as the dynamical
field components of the photon on the light-front in the presence of
magnetic sources. The Hamiltonian density written in terms of these
components is invariant under the duality transformation.
\end{abstract}
\vskip .2in
\centerline{PACS: 12.20.-m, 11.15.-q, 11.10.Ef}
\vskip .2in
\section{Introduction}
\vskip .2in
There has been a resurgence of interest in a new aspect of 
light-front theory in recent
years with the advent of Matrix Theory\cite{ban} which can be formulated in light front
coordinates \cite{su}. As duality is an important concept and tool in Matrix 
Theory, the
question of whether one can formulate it in light-front coordinates is 
important. In this paper we attempt to formulate electromagnetic duality
for a U(1) gauge field theory in the presence of interactions in $3+1$
dimensions. In a recent paper\cite{mar}, it has been
shown that when the free theory is formulated in light-front coordinates
and in the light-front gauge, there exists a transformation known as 
the Susskind transformation \cite{sus} on the
transverse components of the potential, which gives the electromagnetic
duality transformation. 
However, it has been shown \cite{mar} that the Susskind
transformation doesn't hold for the interacting theory if one tries to
incorporate it within the existing framework of ordinary light-front QED.

For a duality symmetric $interacting$ $
theory$
there should be both electric and magnetic sources present. In this case,
one has to consider a theory which deals with them. There exist several such
theories\cite{bl}. In this paper, we have worked with a well known theory of magnetic
monopoles namely that of Zwanziger\cite{zw}. It employs two potentials.
 We have quantized this
theory on the light front in the light front gauge and shown that, out of
 sixteen phase space degrees of
freedom that one starts with, two transverse field components are the only 
dynamical degrees of freedom. This
is the correct number to describe the electromagnetic theory on the light
front. We have shown that they obey the same commutation relation as 
 that between the  
transverse
components of the vector potential in light front QED. Thus Zwanziger's
theory is the theory of the photon on the light front in the presence of a
magnetic current. We have formulated the  
 duality transformation in terms of these
two dynamical field components. This is a transformation between the
transverse components of two different potentials. If one identifies them as the
two independent field components of the photon on the light front in the
presence of a magnetic current, this
duality transformation has a form similar to the Susskind transformation. 
Thus under this identification, one can interprete it as the
 Susskind transformation for the
interacting theory.
\vskip .2in 
\section{Electromagnetic duality for the free theory in light-front
coordinates}

The light-front coordinates are given by, $x^+ = x^0 + x^3, x^- = x^0 -
x^3$, where $x^+$ is the light-front time and $x^-$ is the longitudinal
coordinate. The inner product of two four-vectors,
$a, b$ is defined as,
\be
a^{\mu} b_{\mu} = {1\over 2} a^+ b^- + {1\over 2} a^- b^+ - a^i b^i
\e
where $ i = 1, 2$.

The Lagrangian density for the free electromagnetic theory is,
\be
 {\cal{L}} = -{1\over 4}F_{\mu\nu} F^{\mu\nu}
\e
where $F^{\mu\nu} = \partial^\mu A^\nu - \p^\nu A^\mu$.

In light-front coordinates and in the light-front gauge $A^+ = 0$,
the Lagrangian 
density can be written as,
\be
{\cal{L}} = {1\over 8}(\p^+ A^-)^2 +{1\over 2} \p^+ A^i \p^-A^i -
{1\over 2}\p^+A^i\p^iA^- - {1\over 4} (\p^iA^j-\p^jA^i)^2
\e
where $\p^+ = 2{\p\over {\p x^-}}$,~$\p^- = 2{\p\over {\p x^+}}$ 
 and $\p^i = -{\p\over {\p x^i}}$.
The momenta conjugate to the fields are constrained,
\be
&&\pi_i = {\p{\cal{L}}\over {\p\p^-A^i}} = {1\over 2} \p^+ A^i \\
&&\pi_- = {{\p{\cal{L}}}\over { \p\p^-A^-}} = 0
\e
It can be shown that, in this gauge $A^-$ is a constrained field and
can be eliminated using the equation of constraint,
\be
(\p^+)^2A^- = 2\p^+\p^iA^i
\e
So there are only two independent degrees of freedom, $A^i$(i=1,2)
 in the theory.

The free electromagnetic theory on the light-front is a constrained theory and
has been quantized using the Dirac procedure \cite{st} or the reduced
phase space 
method \cite{zh}. It can be shown that the two dynamical field components, $A^i$
obey the canonical equal (light-front) time commutator given by,
\be
[A^i(x), A^j(y)]_{x^+ = y^+} = -i{1\over 4}\de^{ij} \ep (x^- - y^-) \de^2(x^{\pp} -
y^{\pp})
\e 
The dual tensor is defined as,
\be
{F}^{d\mu\nu} = {1\over 2} \epsilon^{\mu\nu\lambda\rho}F_{\lambda\rho}
\label{f2}
\e
where $\epsilon^{\mu\nu\lambda\rho}$ is totally antisymmetric, 
$\epsilon^{+-12} = 2$.
It can be shown \cite{mar} that under the following transformation of the transverse
components,
\be
A^j \rightarrow {\tilde A}^j = -\epsilon^{ji}A^i
\label{f1}
\e
the electromagnetic field components undergo the duality transformation,
\be
- F^{d\mu\nu}(A^\perp) = F^{\mu\nu}({\tilde A}^\perp),~~~~
F^{\mu\nu}(A^\perp)= F^{d\mu\nu}({\tilde A}^\perp)
\e
So the transformation given by Eq.(\ref{f1}) known as the  Susskind transformation 
is the duality 
transformation for the free theory.

 Since duality symmetry necessitates the
introduction of a magnetic current for the interacting theory, one has to
work with a theory which deals with it. There are many such
theories \cite{bl}. In this paper,
we work with the theory of Zwanziger,
 in which the
duality transformation can be formulated in terms of the two four potentials
involved. So we  formulate this theory on the light-front  and  show
 that on
quantizing it using the Dirac procedure one gets the right number of
dynamical degrees
of freedom (two) and also the normal light-front commutator between them. The
duality transformation can then be formulated solely in terms of these
dynamical degrees of freedom and it is similar in form to the Susskind
 transformation.

\section{Zwanziger's formulation of electromagnetic theory with electric and
magnetic currents}
The Maxwell equations in the presence of magnetic charges are,
\be
\p_\mu F^{\mu\nu} = j^\nu _e, ~~~~~\p_\mu {F}^{d\mu\nu} = j^\nu_g
\e
where
$\p_\mu j^\mu _e = \p_\mu j^\mu _g = 0$.

Zwanziger showed that \cite{zw} the Maxwell equations can be derived from the
following Lagrangian,
\be
{\cal{L}}=-{1\over 2}n_\alpha(\p \wedge q_a)^{\alpha\mu}n^\beta (\p \wedge
q_a)_{\beta\mu} - {1\over 2}\epsilon_{ab} n_\alpha(\p \wedge
q_a)^{\alpha\mu}n^\beta(\p \wedge q_b)^d _{\beta\mu} - {j_a}^\mu q_{a\mu}
\e
with the identification
\be
F^{\mu\nu} = n^\mu n^\alpha(\p \wedge q_1)_\alpha ^\nu - n^\nu n^\alpha(\p
\wedge q_1)_\alpha ^\mu - \epsilon^{\mu\nu\alpha\beta}n_\alpha n_\gamma(\p
\wedge q_2)^\gamma_\beta
\label{f3}
\e
where $a$ runs over 1 and 2 which correspond to two vector
potentials , and $n^\alpha$ is a fixed vector which is usually chosen to be
spacelike \cite{zw}. $(\p \wedge q_b)^d$ denotes the
dual of $(\p \wedge q_b)$ as in Eq.(\ref{f2}) and the wedge product is defined as,
$(A\wedge B)^{\mu\nu} = A^\mu B^\nu - A^\nu B^\mu.$
This Lagrangian is not manifestly covariant because of the presence of the
fixed vector $n^{\mu}$. Covariance is regained only for quantized values of
the coupling constants $(e_n g_m - e_m g_n)$ where $e_n$ and $g_n$ are
electric and magnetic charges.
Moreover, this theory has two vector potentials and hence has sixteen phase
space degrees of freedom. But the number of independent phase space degrees
of freedom for the gauge field sector of QED is four. So the theory is highly
 constrained. Zwanziger had
quantized it  by introducing
 a gauge fixing term. Later Balachandran
et.al. \cite{bal}
did the quantization using the Dirac procedure by making a specific choice
of the fixed vector $n^{\mu}$
\cite{dir}. 
The duality transformation is given in this theory in terms of the two
potentials and the currents as follows:
\be
q^{\mu}_a \rightarrow \ep_{ab}q^{\mu}_b,~~~~~ j^{\mu}_a \rightarrow \ep_{ab}
j^{\mu}_b
\label{dul}
\e

We now formulate Zwanziger's theory in light-front coordinates.
\section{Zwanziger's theory in light-front coordinates}

 We take
$ n^\mu$ to be spacelike and choose,
\be
n^\mu = (n^+, n^-, n^1, n^2) = (0,0,1,0)\label{f14}
\e
The Lagrangian in light-front coordinates for this choice of $n^\mu$
becomes,
\be
{\cal L} = -{1\over 2} (\p \w q_a)^{1+}(\p \w q_a)^{1-} +&& {1\over 2}(\p \w
q_a)^{12}(\p \w q_a)^{12}+ {1\over 4} \ep_{ab} (\p \w q_a)^{1+}(\p \w
q_b)^{2-}\nonumber\\ &&-{1\over 4} \ep_{ab}(\p \w q_a)^{1-}(\p \w q_b)^{2+} 
-{1\over 4} \ep_{ab} (\p \w q_a)^{12} (\p \w q_b)^{+-} \nonumber\\ 
&&~~~~~~~~~~~~~~~~~~~~~~~~~~~~~~~~~~~~~~~~~~~~- j^\mu_a q_{a\mu}
\e
  We take $j^\mu_a$ to be
currents due to external sources.
The Lagrangian equations of motion are,
\be
&&-\ep_{ab}\p^+(\p \w q_b)^{12} + \p^1(\p \w q_a)^{1+} +\ep_{ab}\p^2(\p \w
q_b)^{1+} = j^+_a \label{f4}\\
&&{1\over 2}\p^+(\p \w q_a)^{1-} + {1\over 2} \p^-(\p \w q_a)^{1+} - \p^2(\p
\w q_a)^{12} = j^1_a \label{f5}\\
&&-{1\over 2}\ep_{ab} \p^1(\p \w q_b)^{+-} + \p^1(\p \w q_a)^{12} = j^2_a
\label{f6}\\
&&\ep_{ab} \p^-(\p \w q_b)^{12} + \p^1(\p \w q_a)^{1-} - \ep_{ab}\p^2(\p \w
q_b)^{1-} = j^-_a \label{f7}
\e
It can be verified easily from Eq.(\ref{f3}) that the $F^{\mu\nu}$ for this choice
of $n^\mu$ becomes,
\be
F^{\mu\nu} = - {\delta}^{\mu 1} (\p \w q_1)^{1 \nu} + {\delta}^{\nu 1} (\p \w
q_1)^{1 \mu} - {\epsilon}^{\mu\nu 1 \beta}(\p \w q_2)^1_{\beta}
\e
 
 Using this expression one can verify that Eqs.(\ref{f4})-(\ref{f7}) are the two sets of Maxwell
equations in light-front coordinates. Since Eq.(\ref{f4}) does not involve a
light-front time derivative $\p^- = 2 {\p\over {\p x^+}}$,
 it gives two constraint equations.

In order to describe electromagnetic theory on the light-front, 
Zwanziger's theory has to have more constraints than the equal-time
case, since the number of phase space variables is two  for
the gauge field sector of QED in  light-front coordinates
 \cite{ko}, instead of four as in the equal time
formulation. To quantize such a highly constrained theory, we use 
 the Dirac procedure in which the Poisson bracket in the
classical theory is replaced by a new object called the Dirac bracket which
gives the commutator in the quantum theory.

  We first calculate the momenta
 canonically conjugate to the
fields $q^\mu_a$, defined as,
\be
\pi_{\mu a} = {\p {\cal L} \over {\p
(\p^- q^\mu_a)}}
\e
Neither of these expressions turn out to have any light-front time
derivatives and hence give rise to what are known as primary
constraints. These are,
\be
&&C_{a1} = \pi_{-a} \label{f8}\\
&&C_{a2} = \pi_{-a} - {1\over 4} \ep_{ba}(\p \w q_b)^{12}\label{f9}\\
&&C_{a3}= \pi_{1a} - {1\over 2} (\p \w q_a)^{1+} - {1\over 4} \ep_{ab} (\p \w
q_b)^{2+} \label{f10}\\
&&C_{a4} = \pi_{2a}+ {1\over 4}\ep_{ba} (\p \w q_b)^{1+} \label{f11}
\e
 For $a=1,2$ there are thus eight primary
constraints. In Dirac's method the constraints are set to zero only after
evaluating all the Poisson brackets. 

The canonical Hamiltonian density is defined as,
\be
{\cal H}_C  = \pi_{\mu a} q^\mu_a - {\cal L}
\e
Where $\pi_{\mu a}$ are obtained from Eqs. (\ref{f8}) - (\ref{f11})
 with the primary constraints
set to zero. The primary  Hamiltonian  is defined as,
\be
H_P = {1\over 2} \int dx^- d^2x^\perp [ {\cal H}_C +
v_{a\alpha} C_{a\alpha}]
\e 
where $v_{a\alpha}$ are eight Lagrange multipliers subject to certain
consistency conditions.

In this case the primary Hamiltonian is given by,
\be
H_P = {1\over 2} \int dx^-d^2x^\perp \{ [ -8(\pi_{+a})^2 - (\p^1\pi_{1a} +
\p^2 \pi_{2a} + \p^+ \pi_{+a} )q^-_a + j^\mu_aq_{\mu a} ] +
v_{a\alpha}C_{a\alpha}\}
\e
The Poisson bracket between two quantities A and B is defined as,
\be
[A(x),B(y)]_P = {1\over 2} \int dz^-d^2z^\perp \Big [ {\p A(x)\over {\p
q^\mu(z)}}{\p B(y)\over {\p \pi_\mu(z)}} - {\p B(y)\over {\p
q^\mu(z)}}{\p A(x)\over {\p \pi_\mu(z)}} \Big ]
\e 
 Evaluating the Poisson brackets of the constraints $C_{a\alpha}$ with the
primary Hamiltonian we find that $\Big[ C_{a1}, H_P \Big]_P$ when required to
be zero give two secondary constraints,
\be
C_{a5} = \p^1 \pi_{1a} + \p^2\pi_{2a} + \p^+\pi_{+a} - {1\over 2} j^+_a
\e  
We also find that the constraints $C_{a1}$ and $C_{a5}$ are first class
constraints since they give vanishing Poisson brackets with all other
constraints including themselves. The remaining ones are second class. 
Requiring the Poisson brackets of the second class constraints with $H_P$
to be zero we get some relations among the Lagrange multipliers
$v_{a\alpha}$
for the second class constraints
from which they can be determined consistently. Poisson
bracket of the secondary constraints $C_{a5}$ with $H_P$ do not give any more
constraints. Thus we find that there are ten constraints in the theory of
which four are first class.

 The presence of first class constraints indicates
that there is gauge freedom in the theory. Since there are four first class
constraints 
one has to introduce four gauge conditions. They are to be chosen  such
 that they give non-zero Poisson brackets with the
first class constraints and at the same time are consistent with the
equations of motion.
We choose the light-front gauge, $q^+_a = 0$. It can be easily checked that
these gauge constraints give non-zero Poisson brackets with the first 
class constraints,
$C_{a5}$. We find that in this gauge, two of the equations of motion given
by Eq.(\ref{f6}) become constraint equations. We take these to be the other two
gauge conditions.   
So we get the gauge constraints,
\be
&&C_{a6} = q^+_a \\
&&C_{a7}= {1\over 2}\ep_{ba} \p^1\p^+ q^-_b + \p^1 ( \p \w q_a)^{12} - j^2_a
\e
$C_{a7}$ give non-zero Poisson bracket with the other first class
constraints, $C_{a1}$. Thus, 
after  the gauge constraints are introduced, the first class constraints
become second class. There are now fourteen second class constraints and the
number of independent degrees of freedom is two, as in ordinary light-front
QED.

This fact can also be observed from the Lagrangian equations of motion.
In the light-front gauge, $q^-_a$ can be eliminated using Eq.(\ref{f6}). As mentioned
earlier, Eq.(\ref{f4}) also represents two constraint equations since they do not
contain any light-front time derivative. Hence out of four field components
$q^i_a$ two can be eliminated using them. Thus there are only two dynamical
degrees of freedom since all the momenta are constrained.

 We now proceed with the Dirac formalism  and construct the $14$ by $14$ constraint matrix, the $ij$th
element of which is given by,
\be
(C)_{ij,ab} = [ C_{ai}, C_{bj} ]_P
\e  
The constraint matrix is given in the appendix A. 
It is non-singular and hence can be inverted. The Dirac
bracket between two quantities A and B is given by,
\be
[A,B]_D = [A,B]_P - {1\over 2} \int dx^- d^2x^{\perp} {1\over 2} \int dy^-
d^2y^{\perp} [A,C_{ai}(x)]_p {C^{-1}}_{ij,ab} (x,y) [C_{bj}(y),B]_P
\e
where $C^{-1} $ is the inverse of the constraint matrix. 

We now calculate the Dirac brackets between the field components, $q^i_a$.
We see that only the constraints $C_{a3}, C_{a4} $ and $C_{a5}$ give non-zero
Poisson brackets with these field components. So 
the block of the inverse matrix that contributes to these Dirac brackets is
given below,
\[ \left( \begin{array}{ccc}
C_{33}^{-1}I & i{\sigma}_2C_{33}^{-1} & 0 \\
-i{\sigma}_2C_{33}^{-1} & C_{33}^{-1}I & i{\sigma}_2
C_{24}^{-1}C_{56}^{-1}C_{26} \\
0 & 
i{\sigma}_2
C_{24}^{-1}C_{56}^{-1}C_{26} & 0 
\end{array} \right) \]
The Dirac bracket between the field components are given by,
\be
[q^i_a(x), q^j_b(y)]_D =&&\de^{i1} \de^{j1}\de_{ab} C^{-1}_{33}(x,y) + \de^{i1}
\de^{j2}\ep_{ab} (C^{-1}_{33})(x,y)) \nonumber \\&&+ \de^{i2} \de^{j1}
\ep_{ab}(-C^{-1}_{33}(x,y)) + \de^{i2} \de^{j2} \de_{ab}(C^{-1}_{33}(x,y))
\nonumber \\&& - \de^{i2} \ep_{ab} \p^j_y (C^{-1}_{24} C^{-1}_{56} C_{26}
(x,y)) - \de^{j2}\ep_{ab} \p^i_x (C^{-1}_{24} C^{-1}_{56} C_{26}
(x,y)) 
\e
From this expression we find,
\be
[q^1_a, q^1_b]_D = {1\over 4} \de_{ab} \ep(x^- - y^-) \de^2 (x^{\pp} -y^{\pp})
\label{f12} = [q^2_a, q^2_b]_D 
\label{f16}
\e

\be
[q^1_a ,q^2_b]_D = -{1\over 4} \ep_{ab} \ep(x^- - y^-) \de^2(x^\perp -
y^\perp) = - [q^2_a ,q^1_b]_D
\label{f17}
\e

Here we have used,
\be
{1\over 2} \int dz^- d^2z^{\pp} f^{-1}(x,z) f(z,y) = \de(x^- -
y^-)\de(x^{\pp}-y^{\pp})
\e
where $f(x,y) $ is a function of $x,y$. This gives,
\be
&&C_{33}^{-1}(x,y) = {1\over 4} \ep(x^- - y^-) \de^2(x^{\pp}-y^{\pp})\\
&&(C_{24} C_{56})^{-1}(x,y)) = {1\over 4} \ep(x^- - y^-)\ep(x^1-y^1)
 \de(x^2-y^2)
\label{inv}
\e 
where we have used the antisymmetry property of the Dirac brackets,
$[q^1_a, q^1_b]_D$ and $[q^2_a, q^2_b]_D $. 

In the quantum theory the Dirac brackets are replaced by $(-i)$ times the
commutator. Then from Eq.(\ref{f12}) we find that both the Dirac brackets 
$ [q^1_a, q^1_b]_D$  and $[q^2_a, q^2_b]_D $ give
the canonical commutator between the transverse field components
$A^i$ in normal light front QED.  We can thus take either $q^1_a$ or
$q^2_a$ as
the two dynamical degrees of freedom and all the others as constrained
variables and Zwanziger's theory expressed in terms of them is the theory of the
photon on the light front in the presence of magnetic sources.  After all the 
Dirac brackets are
evaluated, the constraints can be set to zero. Then the only relevant
bracket is that between the dynamical field components. 
 We take $q^1_a$ , $a = 1,2$ as
the dynamical field components. It is interesting to note that they are the same
components of two different potentials. 
In terms of these the Hamiltonian density 
 can be written as,
\be
{\cal{H}} = -{1\over 2} {[ \ep_{ab} \p^1 q^1_b -\p^2 q^1_a + {1\over {\p^+}}
\ep_{ab} j^+_b]}^2 - j^1_a q^1_a -    
\ep_{ab}j^2_a q^1_b - \ep_{ab}j^2_a {1\over {\p^+}}{1\over {\p^1}} j^+_b 
\e

 The Hamiltonian is invariant under
the duality transformation given by Eq.(\ref{dul}), which can be written solely 
in terms of the
dynamical field components as,
\be
q^1_a \rightarrow \ep_{ab}q^1_b,~~~~~ j^{\mu}_a \rightarrow \ep_{ab}
j^{\mu}_b
\label{dull}
\e 
It is clear that the above duality transformation is similar in form to the
Susskind transformation, however, it is a transformation between the same
component of two different
 potentials, whereas the Susskind transformation for the free
theory is between the two transverse components of a single potential. Since
$q^1_a$ are the only two dynamical field components and they obey the same
commutation relation as the $A^i$s, one can identify them as the two
dynamical field components that describe the photon on the light-front in
the presence of a magnetic current. 
In that sense, the duality transformation can be interpreted as the Susskind
transformation for the interacting theory.

 To summarize, in this paper we have formulated electromagnetic duality on
the light-front in the presence of interactions. We have worked with 
Zwanziger's theory with electric and
magnetic currents in light-front coordinates for a specific choice of
$n^{\mu}$.  We have quantized the theory following
 the Dirac procedure and shown that the number of
independent degrees of freedom is two as it should be for the gauge field in 
 light-front QED and
they obey the canonical light-front QED commutator. The duality
transformation  given in terms of the two dynamical field components can
be interpreted  as the Susskind transformation, provided we identify
these dynamical  components as those which describe a photon on the
light-front in the presence of a magnetic current.

We would like to thank Prof. A. Harindranath for bringing the problem of
electromagnetic duality on the light-front to our attention. We are
grateful to Prof. P. Mitra, Prof. A. Harindranath and also to 
Ananda Dasgupta for
helpful discussions.

\appendix
\section{The Constraint Matrix}

The constraint matrix is given by,

\[ \left( \begin{array}{ccccccc}
0 & 0 & 0 & 0 & 0 & 0 & i{\sigma}_2 C_{17} \\
0 & 0 & C_{23}I & i{\sigma}_2C_{24} & 0 & C_{26}I & 0 \\
0 & C_{23}I & C_{33}I & 0 & 0 & 0 &C_{37}I \\
0 & -i\sigma_2 C_{24} & 0 & 0 & 0 & 0 &C_{47}I \\
0 & 0 & 0 & 0 & 0 & C_{56}I & 0 \\
0 & -C_{26}I & 0 & 0 & C_{56}I & 0 & 0 \\
i{\sigma}_2 C_{17} & 0 & -C_{37}I & -C_{47}I & 0 & 0 & 0 
\end{array}  \right) \]

where we have used $(i {\sigma}_2)_{ab} = \ep_{ab}$ and $I$ is the two by two
identity matrix.
The various  nonzero elements $C_{ij}$ are,

\be
&&C_{17} = -{1\over 2} \p^1_y \p^+_y \delta (x^- - y^-)\de^2(x^\pp -
y^\perp)\\
&&C_{23} = {1\over 2} \p^1_y \delta (x^- - y^-)\de^2(x^\pp - y^\perp)\\
&&C_{24} = -{1\over 2} \p^1_y \delta (x^- - y^-)\de^2(x^\pp - y^\perp)\\
&&C_{26} = -\delta (x^- - y^-)\de^2(x^\pp - y^\perp)\\
&&C_{33} = -\p^+_y \delta (x^- - y^-)\de^2(x^\pp - y^\perp)\\
&&C_{37} =  \p^1_y \p^2_y \delta (x^- - y^-)\de^2(x^\pp -
y^\perp)\\
&&C_{47} = - \p^1_y \p^1_y \delta (x^- - y^-)\de^2(x^\pp -
y^\perp)\\
&&C_{56} =  \p^+_y \delta (x^- - y^-)\de^2(x^\pp -
y^\perp)
\e


\end{document}